\documentclass[fleqn]{article}
\usepackage{espcrc2}
\usepackage{graphicx}
\usepackage{amssymb}

\begin{document}

\bibliographystyle{prsty}

\title{Temperature Dependent Neutron Scattering Cross Sections for 
Polyethylene}

\author{ Roger E. Hill and C.-Y. Liu}

\author{Roger E. Hill\address[LANL]{University of California, Physics Division P-23,
        Los Alamos National Laboratory, \\
        Los Alamos, NM 08545, USA}%
        \thanks{Corresponding author. P-23, MS H803, Los Alamos National Laboratory, Los Alamos, NM 08545, USA. Tel:505-667-8754; fax:505-665-4121; e-mail: rhill@lanl.gov} and
        C.-Y. Liu\addressmark
        }

\date{\today}

%\maketitle

\begin{abstract}
This note presents neutron scattering cross sections for 
polyethylene at 296 K, 77 K and 4 K derived from a new scattering kernel for neutron scattering off of hydrogen in polyethylene. The kernel was developed in ENDF-6 format as a set of S($\alpha,\beta$) tables using the LEAPR module of the NJOY94 code package. The
polyethylene density of states (from 0 to sub eV) adopted to derive the new kernel is presented. We compare our calculated room
temperature total scattering cross sections and double differential 
cross sections at 232 meV at various angles with the available experimental data (at room temperature), and then extrapolate the 
calculations to lower temperatures (77K and 4K). The new temperature dependent
scattering kernel gives a good quantitative fit to the available room temperature data and has a temperature dependence that is qualitatively consistent with thermodynamics.
\end{abstract}

%\makecover
%\tableofcontents
%\listoffigures
%\listoftables

\maketitle

%\addtocounter{page}{2}

\section{INTRODUCTION}
 A prototype solid deuterium (SD$_2$) Ultra Cold Neutron (UCN) source 
has been under development for testing at the Los Alamos Neutron Science Center (LANSCE) since the summer of 1998. This involves the use of a cold ``moderator'' surrounded by a beryllium flux-trap in which is 
embedded a tungsten spallation target. The purpose of the ``moderator''/flux-trap system is to optimize a
flux of cold (20 to 40 K) neutrons that interact with a volume of
SD$_2$ at 4 K to produce UCN through inelastic scattering.  
During the prototype testing phase, in order to avoid safety issues 
associated with the use of liquid deuterium or hydrogen, we have  
chosen polyethylene at 4 K as the cold ``moderator'' material.

 In order to simulate the neutronic behavior of this prototype with Monte Carlo methods using MCNP, it was necessary to
develop a scattering kernel in ENDF-6 format\cite{ENDF} for incoherent neutron scattering from hydrogen in polyethylene at cryogenic temperatures, which are not yet available in the official neutron scattering database. This report describes the development of such a kernel in the form of S($\alpha,\beta$) tables for polyethylene at 296 K, 77 K, and 4 K, temperatures relevant to the
different cooling stages of the experiment.

\section{POLYETHYLENE DENSITY OF STATES}
\begin{figure}
%\centerpicture 9.47 in by 6.49 in (RhoAll scaled 700)
%\begin{center}
\includegraphics[width=3.0 in]{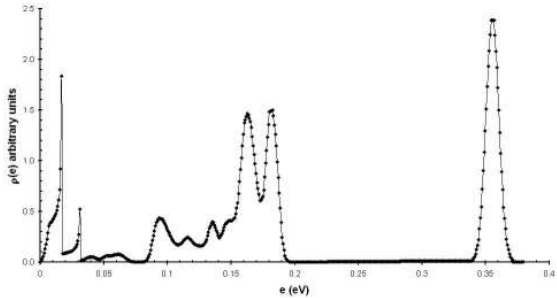}
%\end{center}
\caption{Density of states for polyethylene used for the
scattering kernel.
\label{fig:RhoAll}}
\end{figure}

 A scattering kernel for neutrons scattering off of the
hydrogen in room temperature polyethylene was developed a number of 
years ago by Sprevak and Koppel\cite{Sprevak} using the GASKET\cite{Koppel,Koppel2} code and applying the incoherent approximation .The incoherent approximation is appropriate for polyethylene since the dominant scattering center is the H nucleus where the coherent cross section accounts for only 2.2\% of the total scattering cross section ($\sigma_{tot}$ = 81.67 barn). The density of states of polyethylene used for this scattering kernel calculation was derived from the work of Lin and Koenig\cite{Lin&Koenig}, who calculated the structure force constants assuming polyethylene to be an infinite chain of CH$_2$ radicals (methylene). The force matrix is thus simplified into finite dimensions as a single methylene plus a relative phase parameter between adjacent methylene groups. The density of states (referred to as the frequency distribution in some literature) defined as 
\begin{equation}
\rho (\omega) = \sum_{s} \int \frac{d\vec{k}}{(2\pi)^3} \delta(\omega-\omega_s)
\end{equation}
\noindent can be calculated once the energy spectrum $\omega_s$ is determined. 
Here $s$ indexes energy modes.
The intra-molecular excitations include bond stretching
(symmetric and antisymmetric modes), wagging, bending
and rocking\cite{Tasumi}. They all contribute to the optical modes in the density of states shown in Figure \ref{fig:RhoAll}. Direct excitations of bond stretching results in a sharp peak at 0.36 eV, and the other relative motions between bonds contribute to excitations 
between 0.08 and 0.19 eV. 

\begin{figure}
%\centerpicture 9.47 in by 6.49 in (RhoLow scaled 700)
%\begin{center}
\includegraphics[width=3.0 in]{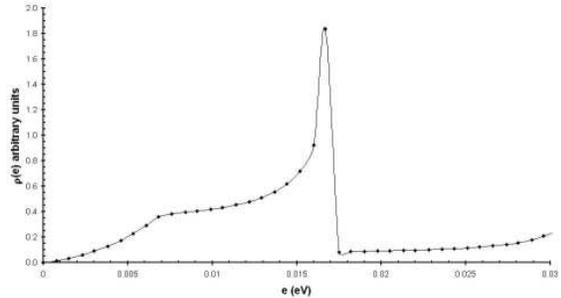}
%\end{center}
\caption{Sub-thermal region of the polyethylene density of states.}
\label{fig:RhoLow}
\end{figure}

 The sub-thermal energy part of the spectrum results from the collective excitations of the solid lattice, which can be approximated by the Debye spectrum\cite{Sprevak} for phonon excitations. A number of years later, Swaminathan and Tewari\cite{Swaminathan}
developed a more detailed polyethylene density of states in the sub-thermal energy range as part of their efforts to explore the use of cold polyethylene as a source of cold neutrons and to
study the effect of the degree of crystallinity on neutron 
scattering. They proposed an anisotropic dispersive continuum model for the acoustic phonons in crystalline polyethylene and a density of states as shown in Figure \ref{fig:RhoLow}. Their paper\cite{Swaminathan} contains a brief description of some of the other scattering kernels that have been developed for polyethylene over the years.

 Our UCN source involves a broad spectrum of neutrons, from the spallation energy of several tens of MeV to a few hundred neV. To perform accurate simulations we require a good knowledge of our neutron moderator, polyethylene, over most of this energy range.
We approach this by combining the epithermal part of the Sprevak and Koppel\cite{Sprevak} spectrum with the low energy part of the Swaminathan and Tewari\cite{Swaminathan} spectrum for crystalline polyethylene. The resulting polyethylene spectrum is shown in Figure \ref{fig:RhoAll}. The low energy ($<$ 30 meV) component of the spectrum, due solely to the acoustic phonons, should be normalized to properly account for the number of degrees of freedom (3 translational for each of the N scatterers(CH$_2$)) in a volume V; i.e.,
\begin{equation}
\int_0^{T_{\Theta}} d\omega \rho(\omega) = 3N/V.
\end{equation}
\noindent  Here $T_{\Theta}$ is the Debye temperature of the polyethylene. This combined spectrum was then
run through the LEAPR\cite{LEAPR} module of the NJOY\cite{NJOY} 
code package to obtain scattering kernels at various temperatures in ENDF-6 format. 

\section{LEAPR FORMALISM}
 Temperature dependent inelastic scattering components can be calculated by integrating the double-differential cross sections obtained from the scattering kernel; i.e. the S($\alpha,\beta$) tables, 
over energy and angle: 
\begin{equation}
\label{eqn:DDS}
  \frac {d^{2}\sigma(T,E_i \rightarrow E_f,\theta )} 
{dE_{f}d\Omega}=\frac{\sigma_b} {4\pi
T}\sqrt{\frac{E_{f}}{E_{i}}} e^{-\beta/2} S(\alpha,\beta), \\
\end{equation}
\noindent where $\sigma_b$ is the bound scattering cross section. 
The unit-less momentum transfer $\alpha$ and energy transfer $\beta$ are
defined as 
\begin{eqnarray}
  \alpha &\equiv& \frac{E_i+E_f-2\sqrt{E_iE_f}\cos\theta}{A T} \nonumber \\
\beta &\equiv& \frac{E_f-E_i}{T}, \nonumber
\end{eqnarray}
\noindent where A is the atomic weight. The scattering kernel $S(\alpha,\beta)$
for inelastic phonon creation/annihilation is 
\begin{equation}
S(\alpha,\beta)=\frac{1}{2\pi}\int_{-\infty}^{\infty}e^{i\beta t}e^{-\gamma(t)}dt
\end{equation} 
\noindent with
\begin{equation}
\gamma(t)=\alpha \int_{-\infty}^{\infty}\frac{\rho(\beta)}{2\beta\sinh(\beta/2)}(1-e^{i\beta t})e^{-\beta/2}d\beta
\end{equation}
\noindent incorporating the material characteristics through its density of
states $\rho(\beta))$.

The incoherent elastic component has a simpler formalism:
\begin{eqnarray}
\label{eqn:DW}
  \sigma(E) = \sigma_b\frac{1-e^{-4EW}}{4EW}
  \mbox{\hspace{0.1 in}  and \hspace{0.1 in}}
  W = \frac{\lambda}{AT}.
\end{eqnarray}
\noindent where the Debye-Waller factor $W$ is determined once the Debye-Waller coefficient $\lambda$ is calculated by LEAPR.

\section{COMPARISON OF THE SCATTERING KERNEL WITH EXPERIMENTAL DATA.}
\begin{figure}
%\centerpicture 9.47 in by 6.49 in (SigTot scaled 700)
%\begin{center}
\includegraphics[width=3.5 in]{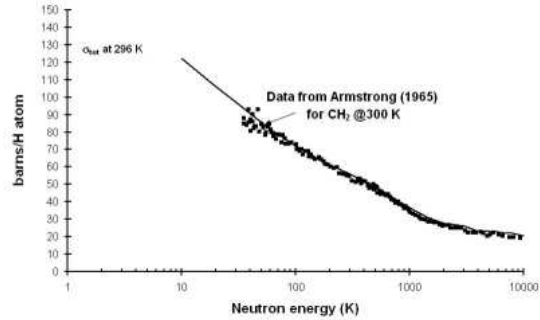}
%\end{center}
\caption{Total neutron scattering cross section from room temperature 
polyethylene, calculated using the density of states presented in Fig.\ref{fig:RhoAll}. The result of calculation (solid curve) is compared with experimental data (dots).}
\label{fig:SigTot} 
\end{figure}

 We have tested the proposed polyethylene density of states by using it to generate room temperature cross sections in LEAPR and comparing the results with available data. A comparison of the calculated and measured\cite{Amstrong} total scattering cross section
for room temperature polyethylene is shown in Figure \ref{fig:SigTot}. The experimental data have been adjusted by subtracting a 1/v hydrogen absorption cross section (equal to 0.334 b/H atom at 0.0253 eV) and a constant 4.74 barns for scattering from the carbon nucleus. The agreement between the theory and the measurements is excellent. 

\begin{figure}
%\centerpicture 9.47 in by 6.49 in (dds_25 scaled 700)
%\begin{center}
\includegraphics[width=3.3 in]{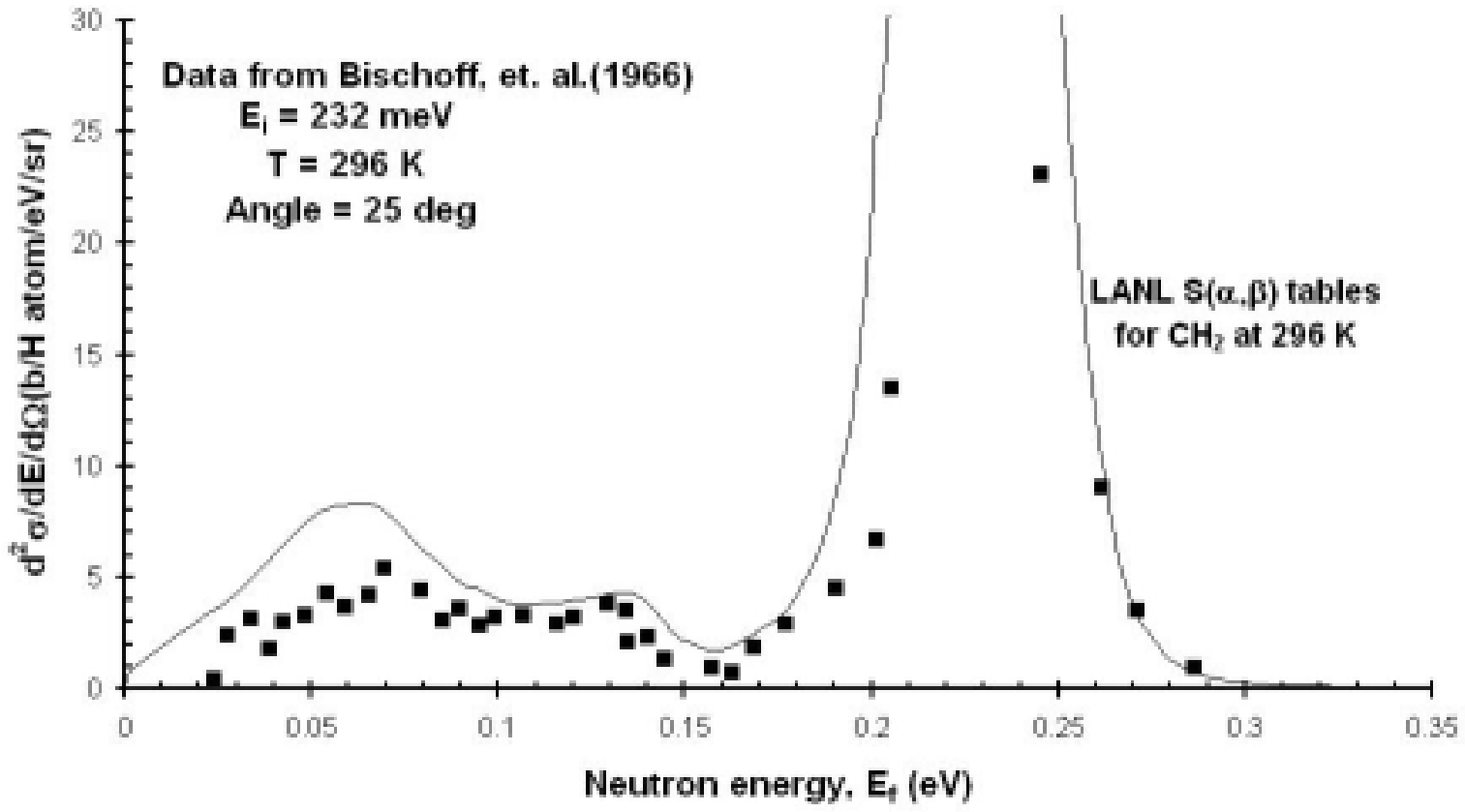}
\includegraphics[width=3.3 in]{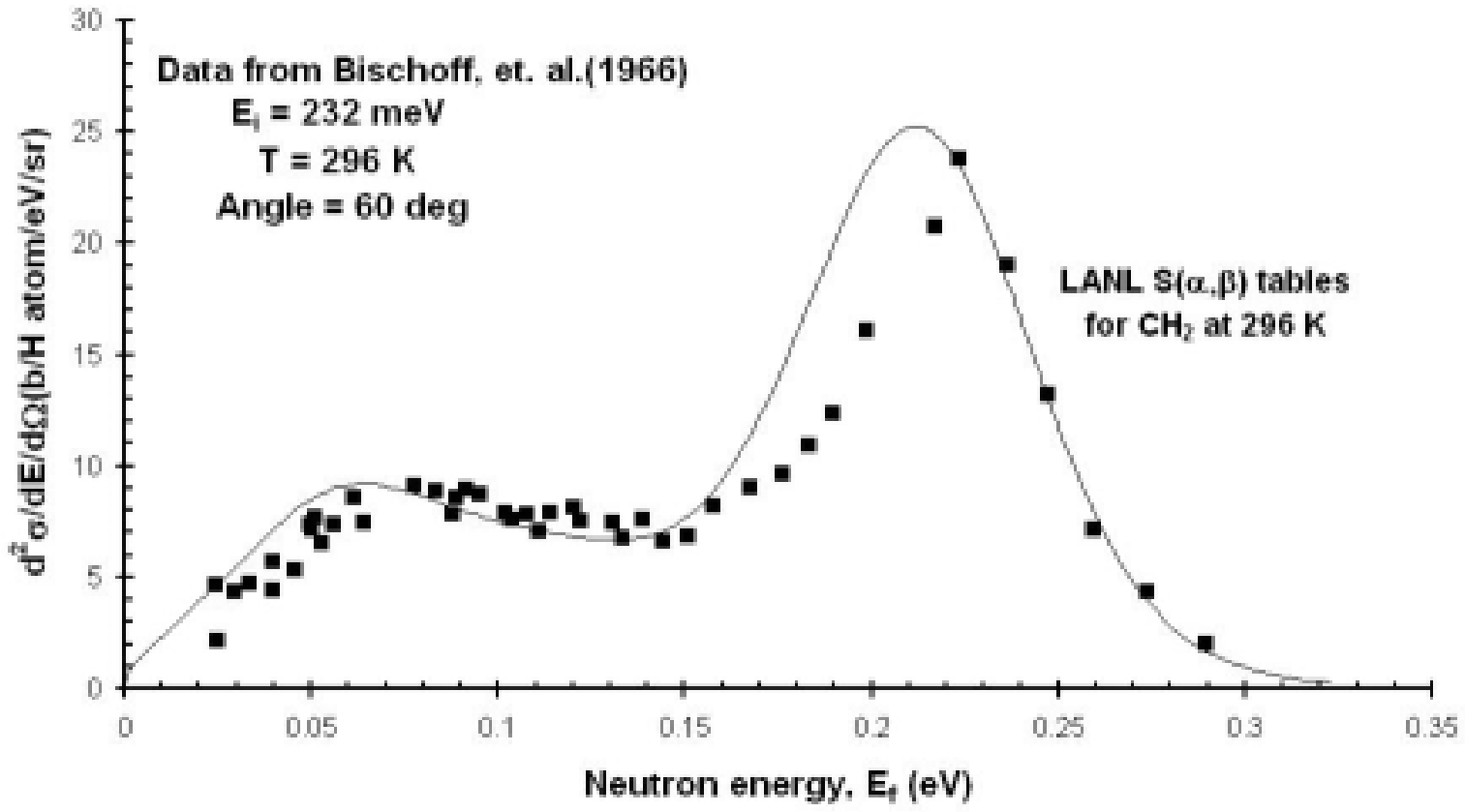}
\includegraphics[width=3.3 in]{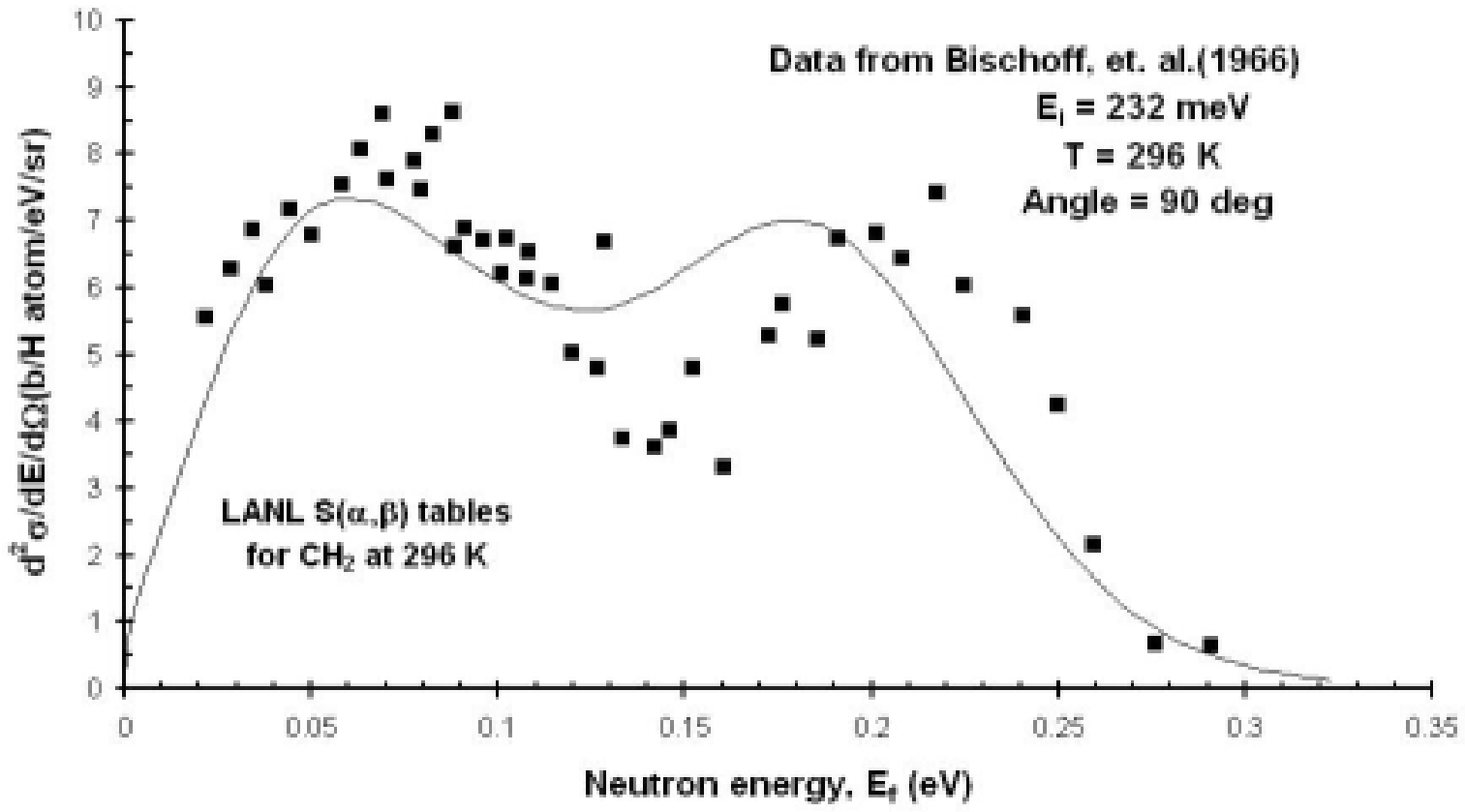}
%\end{center}
\caption{Double differential cross sections for 232 meV incident neutron at 25, 60 and 90 degree scattering off of room temperature (296K) polyethylene. 
The results of calculation (solid curves) are compared with experimental data (dots).
\label{fig:dds_25} }
\end{figure}

 The polyethylene density of states is further tested by looking into the details of the double-differential cross-sections obtained from LEAPR. A direct comparison of these double-differential cross sections with experimental data\cite{Bischoff} at three different angles for an incident neutron energy of 232 meV
(T$_n$ = 2700 K) is given in Figure \ref{fig:dds_25}.

The experimental data were taken from figures in the Sprevak and
Koppel paper\cite{Sprevak}. Once again, the agreement is satisfactory. Overall, the structure built into the S($\alpha,\beta$)
tables from the $\rho(e)$ matches well the structure observed in the neutron 
scattering. The peaks at low energy correspond to the excitations of 
the energy states around 0.16 eV and 0.095 eV 
(seen in Figure \ref{fig:RhoAll}). In addition, there is little 
up-scattering (no energy-gain peaks). 
This is what we would expect in situations 
of predominately inelastic scattering where the incident neutron 
energy is much larger than the moderator energy (T$_n >>$ T). 

\section{EFFECT OF POLYETHYLENE TEMPERATURE ON COLD NEUTRON SCATTERING.}

\begin{figure}
%\centerpicture 9.47 in by 6.49 in (elastic scaled 700)
%\begin{center}
\includegraphics[width=3.5in]{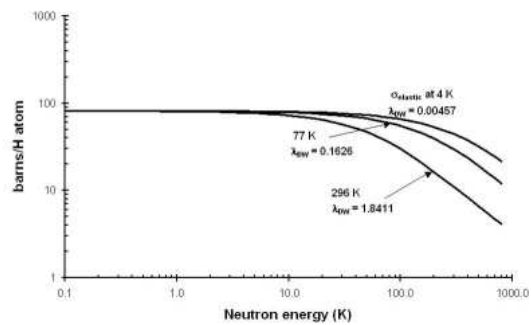}
%\end{center}
\caption{Calculated neutron elastic scattering cross section for
polyethylene temperatures
of 296, 77 and 4 K. Also shown are the Debye-Waller coefficients at
each temperature.}
\label{fig:elastic}
\end{figure}

Equation \ref{eqn:DW} implies that the elastic scattering cross section is temperature independent in the limit of low neutron energy and inversely dependent on temperature in the high energy limit. Plotted in Figure \ref{fig:elastic} is the effect of polyethylene temperature on incoherent elastic scattering. As expected, the elastic scattering of cold neutrons (T$_n<$ 10K, or 1 meV) is independent of polyethylene temperature, and the cross section at high energies scales roughly as 1/T.

\begin{figure}
%\centerpicture 9.47 in by 6.49 in (inelastic scaled 700)
%\begin{center}
\includegraphics[width=3.5in]{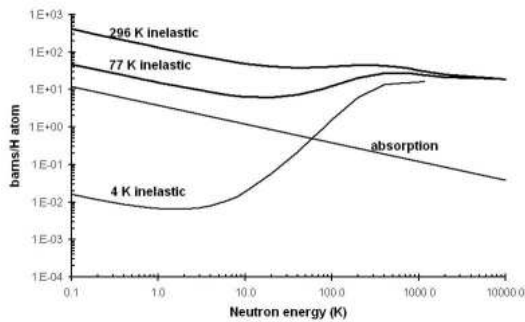}
%\end{center}
\caption{Calculated neutron inelastic scattering cross section for polyethylene
temperatures of 296, 77 and 4 K, compared with the nuclear absorption cross section.}
\label{fig:inelastic}
\end{figure}

 For incoherent inelastic scattering, on the other hand, Figure \ref{fig:inelastic} shows that low energy neutron scattering is strongly dependent on polyethylene temperatures while epithermal scattering (T$_n > $1000 K or 100 meV) is almost temperature independent. In the high energy region (relative to the moderator temperature) the dominant process is down-scattering through excitation of internal states of the material. These states are intrinsic and their availability depends very little on temperature. For neutrons at very low energies, on the other hand, the thermalization process requires up-scattering where the cross section scales as the number of available heat carriers; i.e., the phonon density. The phonon number density is governed by the Boltzmann factor and can be suppressed by cooling the material. These solid state effects produce the observed temperature dependence.

Moreover, it should be noted that the usefulness of CH$_2$ at 4 K as a moderator
is limited by nuclear absorption. The absorption cross section scales inversely with the neutron velocity. For CH$_2$ at 40 K, inelastic scattering has a larger amplitude than absorption at all neutron energies. For CH$_2$ at 4 K, on
the other hand, the nuclear absorption overwhelms the inelastic scattering for all neutrons with energies less than 70 K, resulting in a smaller output flux of cold neutrons. For the LANL UCN source, where substantial quantities of cold neutrons with energies around 40 K are desired (these are subsequently cooled to UCN energies by inelastic scattering in solid D$_2$), the absorption losses are kept acceptable by using 4 K CH$_2$ thicknesses in the 10-20 mm range. 

\begin{figure}
%\centerpicture 9.47 in by 6.49 in (cold_296 scaled 700)
%\begin{center}
\includegraphics[width=3.3in]{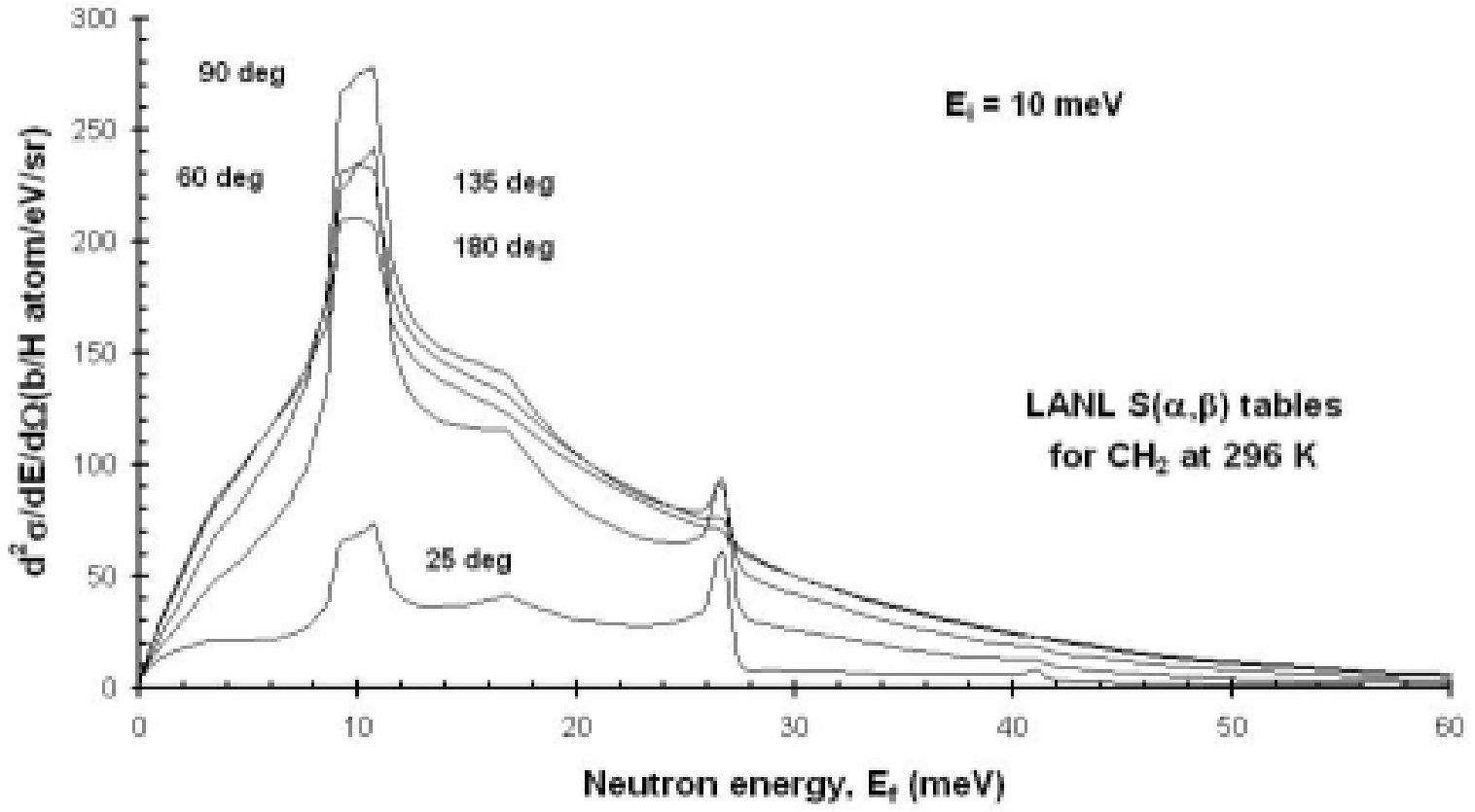}
\includegraphics[width=3.3in]{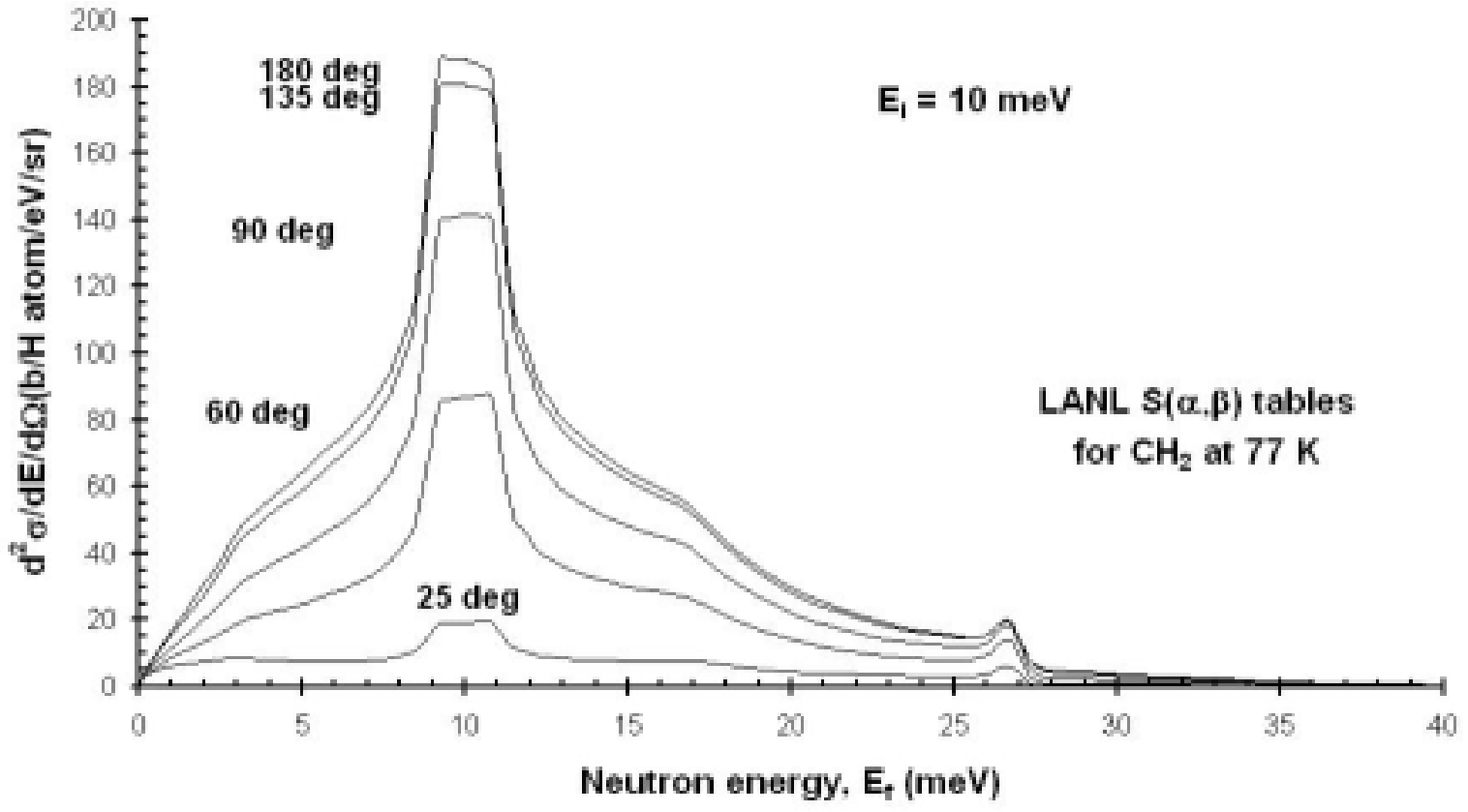}
\includegraphics[width=3.3in]{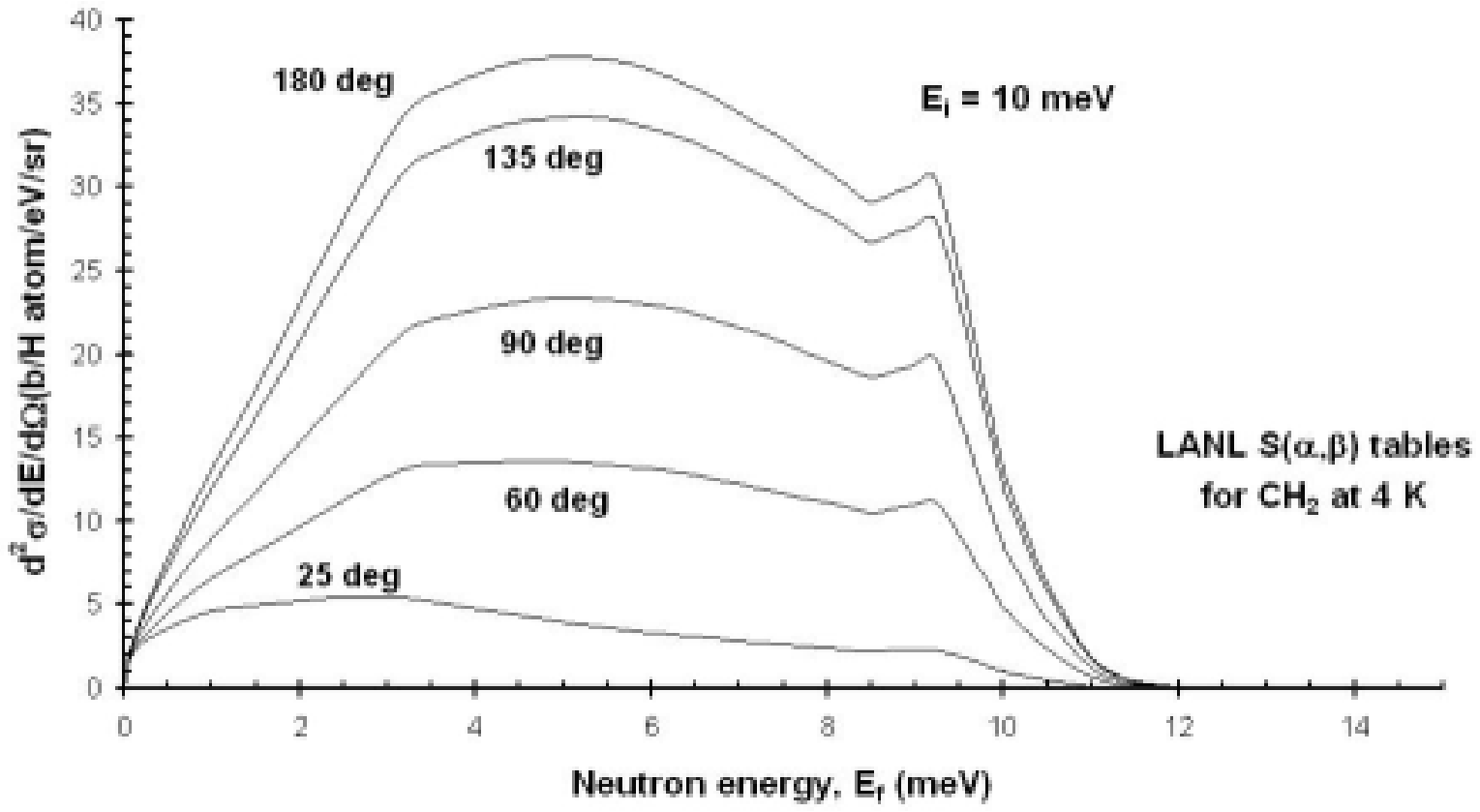}
%\end{center}
\caption{Calculated differential cross section for scattering of a 10
meV (116 K) neutron
from polyethylene at 296 K, 77K and 4K.}
\label{fig:cold_296}
\end{figure}

  Finally, we use Figure \ref{fig:cold_296} to illustrate the dynamics of neutron scattering off of moderators at different temperatures. Figure \ref{fig:cold_296} presents the differential scattering cross sections for a 10 meV (116 K) neutron from polyethylene at 296, 77 and 4 K. At 296 K, the incident neutrons with E$_i$ = 10 meV have energies lower than the average thermal energy of the moderator and thus most of the scattering amplitude is on the energy gain side (up-scattering). In addition, the scattering amplitude is almost uniform in angle; the peak at 90 degree merely represents the cos$\theta$d$\theta$ weighting of the solid angle. This 4$\pi$ scattering is consistent with the random direction of phonon propagation in polyethylene.
As the moderator temperature lowers, more down-scattering occurs and the angular distribution becomes backward peaked for maximum energy transfer to phonons. 
At 77 K the up and down scattering are 
nearly equal. At 4 K, there is 
almost no up-scattering and the
down-scattering is backward peaked. This variation in  up and
down-scattering is to be expected for neutron-moderator systems tending towards thermal equilibrium.

\section{CONCLUSIONS.}
 The new temperature dependent scattering kernel for polyethylene gives a good quantitative fit to the neutron scattering data available at room temperature and has a temperature dependence that is qualitatively consistent with thermodynamics. The kernels are available upon request from rhill@lanl.gov or 
cyliu@lanl.gov.

%\section{Acknowledgments}

%\newpage
%\ \ \\
%\vspace{3.0in}

%\begin{center}
%This page has been intentionally left blank.
%\end{center}

\end{document}